\documentclass[prd, preprint, nofootinbib]{revtex4}
\usepackage{graphicx}
\usepackage{amsfonts}
\usepackage{latexsym}
\usepackage{amssymb}
\usepackage{epsfig}

\begin{document}

\title{ Axion monodromy inflation with sinusoidal corrections
}

\author{Tatsuo Kobayashi$^1$, Osamu Seto$^2$, and Yuya Yamaguchi$^1$}
 \affiliation{$^1$Department of Physics, Hokkaido University, Sapporo 060-0810, Japan \\
$^2$Department of Life Science and Technology,
  Hokkai-Gakuen University, Sapporo 062-8605, Japan
}

%
\begin{abstract}
We study the axion monodromy inflation with a non-perturbatively 
generated sinusoidal term.
The potential form is a mixture between the natural inflation 
and the axion monodromy inflation potentials.
The sinusoidal term is subdominant in the potential, but 
 leaves significant effects on the resultant fluctuation generated
 during inflation.
A larger tensor-to-scalar ratio can be obtained in our model.
We study two scenarios, the single inflation scenario and 
the double inflation scenario.
In the first scenario,
 the axion monodromy inflation
 with a sufficient number of e-folds 
 generates a larger tensor-to-scalar ratio of about 0.1-0.15
 but also a tiny running of the spectral index.
In the second scenario of double inflation, 
 axion monodromy inflation is its first stage,
 and we assume another inflation follows.
In this case, our model can realize a larger tensor-to-scalar ratio
 and a large negative running of the spectral index simultaneously.
\end{abstract}

\pacs{}
\preprint{EPHOU-14-009}
\preprint{HGU-CAP-033}

\vspace*{3cm}
\maketitle


\section{Introduction}

Recent detection of the gravitational wave perturbation
 by BICEP2~\cite{Ade:2014xna} indicates its large amplitude
  and its tensor-to-scalar ratio of
\begin{equation}\label{eq:BICEP2-1}
r _T = 0.20^{+0.07}_{-0.05}
\end{equation}
 for a lensed-$\Lambda$CDM plus tensor mode cosmological model, and
\begin{equation}\label{eq:BICEP2-2}
r _T = 0.16^{+0.06}_{-0.05}
\end{equation}
 after foreground subtraction based on dust models.
This implies that the energy scale of inflation is high and
 the inflation potential would belong to the so-called 
 large field model, where an inflaton takes a super-Planckian field value,
 such as chaotic inflation~\cite{Linde:1983gd,ChaoticAfterBicep}.

However, it is non-trivial to control a flat potential 
with a super-Planckian field value.
An axion is one of the interesting candidates for inflaton fields, 
 because it has a shift symmetry and its potential would be 
 flat for a super-Planckian field value.
In this sense, the so-called natural inflation~\cite{Freese:1990rb} is interesting,
 and when studied in light of the BICEP2 data~\cite{NaturalAfterBicep}.
Furthermore, axion monodromy inflation 
\cite{Silverstein:2008sg,McAllister:2008hb,Flauger:2009ab}
 has been proposed as an axion inflation model
 within the framework of low-energy effective 
 field theory derived from superstring theory 
(see also \cite{Kaloper:2008fb}).\footnote{
For a review, see \cite{Baumann:2014nda}, and also for recent works \cite{recentwork}.}
While in natural inflation the potential form is 
 a sinusoidal function,
 in a simple axion monodromy model the potential is a linear function of 
 the inflaton, $V=a\phi$.\footnote{
The potential $V=a \phi^r$ with $r$ being a fractional number is 
 also possible. In addition, such a potential can be also 
derived by field-theoretical strong dynamics~\cite{Harigaya:2012pg}
 or by non-canonical kinetic terms~\cite{HMLee}.}

Axion monodromy inflation with the linear potential
 is compatible with the data.
However, if we examine it in detail, there are a few issues.
One is that the linear potential inflation predicts
 the scalar spectral index $n_s \sim 0.96$ as indicated 
 by PLANCK~\cite{Ade:2013zuv} and $r_T \sim 0.08$.
The predicted $r_T$ looks a bit too small compared with BICEP2 results
 even if we compare with Eq.~(\ref{eq:BICEP2-2}).
Another is a tension between BICEP2 and PLANCK,
 which is a problem not only for this particular model but for most inflation models.
BICEP2 reported a somewhat larger tensor-to-scalar ratio
 than the upper bound $r_T  < 0.11$ obtained by PLANCK.
As is pointed out in Ref.~\cite{Ade:2014xna},
 the tension can be relaxed
 if the running of the scalar spectral index, $\alpha_s$, is negatively large as 
 $\alpha_s \sim -$(0.02-0.03) \cite{Ade:2013zuv} (see also \cite{Li:2014cka}).

In this paper, we study the inflation model driven by 
 the linear potential derived from the axion monodromy inflation model,
 taking a correction term for non-perturbative effects into account.
That is, non-perturbative effects would generate a correction term for the axion 
in the potential, whose form is the sinusoidal function.
Such a correction was mentioned in Ref.~\cite{McAllister:2008hb}, 
 but it has been neglected because it is subdominant in the potential.
However, we point out non-negligible contributions by the correction term  
 in derivatives of the potential.\footnote{
See also \cite{Kobayashi:2010pz},
 where the same potential was used to explain WMAP seven-year data with $n_s \sim 1.03$ and $\alpha_s \sim -0.03$
 without considering tensor-to-scalar ratio.}
Since the slow roll parameters are expressed
 in terms of the derivatives of the potential,
 the resultant correction in the slow roll parameters
 is quite important in the evaluation of the inflationary observables.
Thus, we study the inflation potential with the linear term and sinusoidal term.

Our purpose is two-fold, and we study two scenarios for each of these.
First, we point out that the sinusoidal correction term increases the predicted $r_T$
 to as much as 0.1-0.15 for the number of e-folds $N=$ 50-60, which
 corresponds to the cosmological scale $k_* = 0.002\, {\rm Mpc}^{-1}$.
This size of $r_T$ is within the range of Eq.~(\ref{eq:BICEP2-2})
 as reported by BICEP2.
We estimate observables $n_s$, $r_T$, and $\alpha_s$ for several parameter sets
 in our model.
Next, we study the case where $r_T \simeq$ 0.1-0.2 and
 the tension between BICEP2 and PLANCK is resolved
 by a large negative running $\alpha_s$.
The sinusoidal correction term can induce 
 negative running by its potential derivatives; 
 however, the size is not enough as long as we assume
 that both the horizon and the flatness problem are solved
 only by this axion monodromy inflation.
On the other hand, if we regard this monodromy inflation as the first stage 
 in the double inflation scenario~\cite{DoubleInflation}
 followed by another inflation,
 we can obtain favored values of the observables
 $(n_s,r_T,\alpha_s) \sim (0.96, 0.15,-0.02)$.
In a previous work,
 this kind of possibility was studied by two of the present authors (T. K. and O. S.)
 in the context of the supersymmetric hybrid inflation model~\cite{Kobayashi:2014rla}.\footnote{
A large $r_T$ from double inflation was also studied in Ref.~\cite{Choi:2014aca}.}

Our paper is organized as follows.
In Sect. II, we explain our model explicitly.
In Sect. III, we study observables derived from our model 
in both the single and double inflation scenarios.
Section IV is devoted to conclusion and discussion.

\section{``Natural'' monodromy inflation}

Before showing our model explicitly, 
we note the definition of several quantities and show 
several formulas.
We use units with $8 \pi G =1$.
The slow-roll parameters for 
 the inflation potential $V$ of the inflaton $\varphi$ are 
 defined as 
 \begin{eqnarray}
 \eta &=& \frac{V_{\varphi\varphi}}{V} , \\
 \varepsilon &=& \frac{1}{2}\left(\frac{V_{\varphi}}{V}\right)^2 , \\
 \xi &=& \frac{V_{\varphi}V_{\varphi\varphi\varphi}}{V^2} ,
\end{eqnarray}
where $V_{\varphi}$ denotes the first derivative of $V$, 
and $V_{\varphi\varphi}$ and $V_{\varphi\varphi\varphi}$ have similar meanings.
The power spectrum of the density perturbation is given by
\begin{eqnarray}
{\cal P}_{\zeta} &=& \left(\frac{H^2}{2\pi |\dot{\varphi}|}\right)^2
 = \frac{V}{24 \pi^2 \varepsilon} .
\end{eqnarray}
In the following analysis, a parameter which represents the height of the scalar potential is
 normalized by this with the amplitude of the temperature anisotropy of
 the cosmic microwave background radiation.
The scalar spectral index, its running and the tensor-to-scalar ratio are expressed as
\begin{eqnarray}
n_s &=& 1 + 2 \eta -6 \varepsilon ,\label{Formula:ns}\\
\alpha_s &=& 16 \varepsilon\eta -24 \varepsilon^2 -2\xi ,\label{Formula:alphas}\\
r_T &=& 16 \varepsilon ,\label{Formula:rT}
\end{eqnarray}
 respectively by use of the above slow-roll parameters.
Also the number of e-folds is evaluated as 
\begin{eqnarray}
N = \int^{\varphi}_{\varphi_e} \frac{V}{V_{\varphi}} d\varphi,
\end{eqnarray}
where $\varphi$ at the upper bound of the integral is the field value 
 when the number of e-folds is $N$ counting from the end of inflation,
and
 $\varphi_e$ denotes the field value at the end time of inflation.

For $n_s \sim 0.96$ and $r_T \sim 0.16$, we find
 $\varepsilon \sim 0.01$ and $\eta \sim 0.01$ from
 Eqs.~(\ref{Formula:ns}) and (\ref{Formula:rT}).
Then, with vanishing $\xi$, we expect $\alpha_s = {\cal O}(0.001)$.
Thus, $\xi$ is important to realize a large negative value 
of $\alpha_s$, and $\xi = {\cal O}(0.01)$ is required 
for $|\alpha_s|={\cal O}(0.01)$.

Now we write the inflation potential of our model,
\begin{eqnarray}
V = a_1 \phi + a_2 \cos \left( \frac{\phi}{f} + \delta \right) + v_0,
\end{eqnarray}
where $a_1$, $a_2$, $f$, $\delta$ and $v_0$ are constants.
The first linear term of $\phi$ would appear 
in the large $\phi$ limit from the term
 $\mu \sqrt{1+(\phi/M)^2}$, which could originate from the Dirac-Born-Infeld action 
 \cite{Silverstein:2008sg,McAllister:2008hb}.
We assume that the second term is generated by non-perturbative effects.
Indeed, this was also mentioned in Ref.~\cite{McAllister:2008hb}, 
but it has been neglected because it may be subdominant.
However, as we will show, even if it is subdominant in $V$, it would be 
important in the derivatives, in particular the second and third 
derivatives of the potential.
Thus, we consider both the first and second term.
We have added the constant $v_0$ such that $V=0$ at the potential minimum.
However, when the second term is subdominant in the potential, 
 we can see that the constant $v_0$ is neglected.
Using this potential, we will study the single and double inflation scenarios in the next section.

\section{Results}

\subsection{Single inflation scenario}

Here, we consider the case that
 the axion monodromy inflation expands enough
 to solve the horizon and flatness problem and study other observables.
Let us call this case the single inflation scenario,
 where the cosmological scale corresponds to $N=$ 50-60.  
First of all, the limit with $a_2 \rightarrow 0$ is quite simple.
We obtain 
\begin{eqnarray}
\varepsilon = \frac{1}{2\phi^2}, \qquad \eta = \xi = 0,
\end{eqnarray}
and 
\begin{eqnarray}
N = \frac{1}{2} (\phi^2 - \phi_e^2).
\end{eqnarray}
Then, when $\phi \sim 10$, we estimate 
\begin{eqnarray}
N \sim 50, \qquad n_s \sim 0.97, \qquad r_T \sim 0.08, 
\end{eqnarray}
which are consistent with PLANCK and BICEP2
 (see, e.g., \cite{Harigaya:2014sua}).
In addition, we estimate the running as $\alpha_s \sim -0.0006$.
A large negative value of $\alpha_s$ is not realized in this case.
That is obvious because in this case we can write 
\begin{eqnarray}
n_s -1 = -\frac{3}{2N}, \qquad r_T = \frac{4}{N}, 
\qquad \alpha_s = -\frac{3}{2 N^2}.
\end{eqnarray}
We would have a large negative $\alpha_s$ for smaller $N$.
However, in this case, $n_s$ becomes too small.

\begin{figure}[!t]
\begin{center}
\epsfig{file=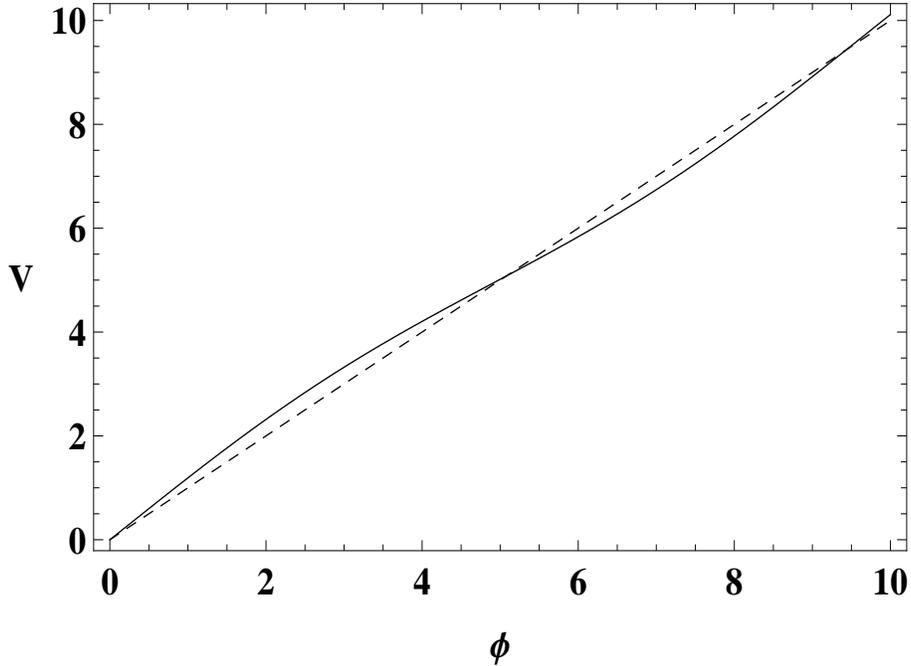, width=12cm,height=9cm,angle=0}
\end{center}
\caption{ The shape of the scalar potential where the vertical axis is in an arbitrary unit and 
 rescaled by $a_1$. For this figure, we take $A=0.3$, $\delta=-1.7$, and $x=0.2$. 
The solid and dashed lines represent with and without the sinusoidal correction, respectively.
 }
\label{Fig:potential}
\end{figure}

Next, we consider the potential with non-vanishing $a_2$.
Then, the slow-roll parameters can be written as 
\begin{eqnarray}
\varepsilon &=& \frac{1}{2\phi^2}\left( \frac{1-x \sin \theta}{1+A \phi^{-1}
\cos \theta }\right)^2, \\
\eta &=& - \frac{x^2}{A \phi} \frac{\cos \theta}{(1 + A \phi^{-1}\cos \theta )} , \\
\xi &=& -\frac{x}{A} \sqrt{2\varepsilon} \eta \tan \theta,
\end{eqnarray}
where $A = a_2/a_1$, $\theta = \phi/f + \delta$, and $x= A/f$.
We show the modification of axion monodromy inflation by including the non-perturbative correction of the sinusoidal term in Fig.~\ref{Fig:potential}. 
%
%

\begin{figure}[!t]
\begin{center}
\epsfig{file=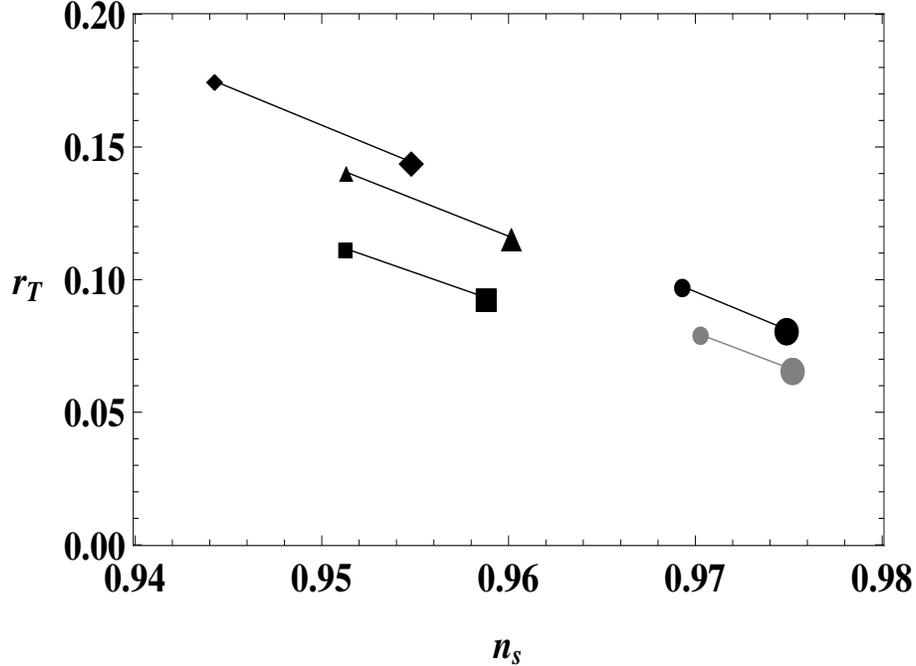, width=12cm,height=9cm,angle=0}
\end{center}
\caption{ The predicted $n_s$ and $r_T$ for $N=$ 50-60
 by some of benchmark parameter sets
 in Table~\ref{tab:single}. The gray corresponds to the linear potential.
The circle, square, triangle, and diamond denote $n_s$ and $r_T$ using $x$, $A$, and $\theta$ of S1, S2, S4, and S7, respectively.
 Smaller and larger symbols correspond to $N=50$ and $60$, respectively.
 }
\label{Fig:nsrt}
\end{figure}

For large $\phi$, we could approximate 
\begin{eqnarray}
\varepsilon \approx \frac{1}{2\phi^2}\left( {1-x \sin \theta}\right)^2, \qquad 
\eta \approx - \frac{x^2 \cos \theta}{A \phi} .
\end{eqnarray}
For example, when $-x \sin \theta = 0.4$, we obtain $( {1-x \sin \theta})^2 \approx 2$.
Then, we have a factor of $2$ enhancement on $\varepsilon$,
 which results in that on $r_T$ as well,
 for a fixed value of $\phi$.
Even for $-x \sin \theta = 0.1$, we would have significant shifts on 
$\varepsilon$ and $r_T$.
On the other hand, as $\varepsilon$ becomes larger, 
$n_s$ becomes smaller than 0.96.
We have to cancel such a shift of $n_s$ by increasing $\eta$ 
to keep $n_s \simeq 0.96$.
Thus, the case with $\cos \theta =0$ is not realistic, 
but both $\cos \theta$ and $\sin \theta$ must be of ${\cal O}(0.1)$-${\cal O}( 1)$.
It is also favorable that both $\cos \theta $ and $\sin \theta $ are negative.
In this case, $\xi$ is always negative, thus
 $\alpha_s$ are mostly positive and can be only tiny negative.\footnote{
On the other hand, we start our discussion to derive a large negative $\alpha_s$ at first.
Because of $\xi \simeq \frac{\sqrt{2 \varepsilon} x^3 }{A^2 \phi} \sin \theta$, 
 we are required to obtain $\sin \theta >0$.
In this case, we have $(1- x \sin \theta)^2 <1$ in Eq.(19).
Thus, we obtain smaller $r_T$ than the linear potential without the sinusoidal term, i.e. $r_T < 0.08$.
Here, we concentrate on the linear potential with the sinusoidal correction.
If we consider the potential term $\phi^p$ instead of the linear term, 
 the situation would change.
For $p \gtrsim 2$, we already obtain $r_T \gtrsim 0.16$ without the sinusoidal term.
Again we need $\sin \theta >0$ to realize a large negative $\alpha_s$
 and it has a negative correction on $r_T$.
However, even including such a negative correction, we can realize $r_T \sim 0.1$ for $p \gtrsim 2$.}
From the above reasoning, it is impossible to realize
 $N \sim 50$, $n_s \sim 0.96$, $r_T \sim 0.1$ and a large negative $\alpha_s$ simultaneously.
We show some examples in Table~\ref{tab:single} and Fig.~\ref{Fig:nsrt}, 
where we fix $\phi = 10$ and $\delta$ can be 
obtained as $\delta = \theta - \phi/f$.

\begin{table}[h]
\begin{tabular}{|c|cccc|cccc|}\hline
&~~~~$x$ ~~~~&~~~~  $A$ ~~~~&~~~~ $\phi$ ~~~~&~~~~ $\theta$ ~~~~
& ~~~~$n_s$ ~~~~&~~~~ $r_T$~~~~ &~~~~ $N$ ~~~~&~~~~ $\alpha_s$~~~~ \\
\hline \hline
S1& 0.1 & 0.2 & 10 & $47 \pi / 36$ & 0.969 & 0.098 & 50 & $-0.00015$ \\
S2& 0.2 & 0.3 & 10 & $19 \pi / 12$ & 0.951 & 0.11 & 50 & 0.00046 \\
S3&0.2 & 0.4 & 10 & $23 \pi / 18$ & 0.969 & 0.12 & 51 & 0.00048 \\
S4&0.3 & 0.4 & 10 & $53 \pi / 36$ & 0.957 & 0.13 & 55  & 0.0028 \\
S5&0.3 & 0.5 & 10 & $53 \pi / 36$  & 0.951 & 0.14 & 51 & 0.0013 \\
S6&0.3 & 0.6 & 10 & $47 \pi / 36$ & 0.965 & 0.14 & 51 & 0.0012 \\
S7&0.4 & 0.6 & 10 & $13 \pi / 9$ & 0.953 & 0.15 & 58 & 0.0033 \\
S8&0.4 & 0.7 & 10 & $25 \pi / 18$   & 0.955 & 0.16 & 55 & 0.0025 \\
\hline
\end{tabular}
\caption{Results in the single inflation scenario}
\label{tab:single}
\end{table}

%
%

\subsection{Double inflation scenario}

Here, we study the double inflation scenario, where 
 the present cosmological scale corresponds
 to a smaller number of $N$
 in the first monodromy inflation and 
 the second inflation derives another
 large e-fold of $N\sim$ 30-50 to solve
 the horizon and flatness problems.
The second inflation can be any of the small field inflation models, 
 if it generates enough e-folding.

\begin{figure}[!t]
\begin{center}
\epsfig{file=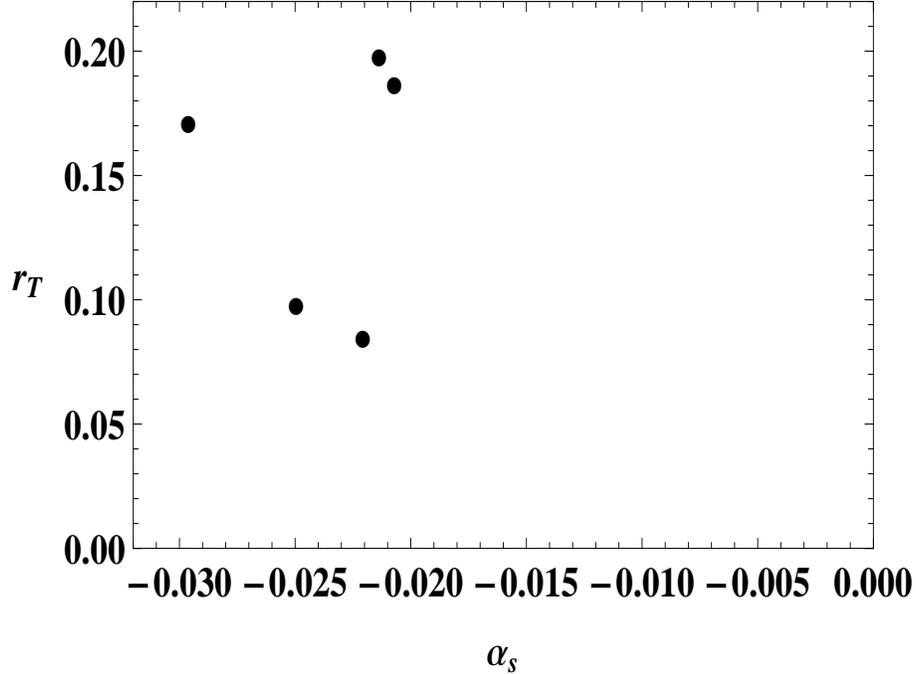, width=12cm,height=9cm,angle=0}
\end{center}
\caption{ The predicted $\alpha_s$ and $r_T$
 by the benchmark parameter sets
 in Table~\ref{tab:double}. 
 }
\label{Fig:asrt}
\end{figure}

In this scenario, larger $\varepsilon$, and as a result larger $r_T$,
 can be obtained because 
 the field value at the $k_*$ mode horizon crossing, $\phi_*$ is smaller.
Since the enhancement of $\varepsilon$ can be done with small $\phi$,
 $\sin \theta$ can be positive or negative.
To keep $n_s \simeq 0.96$ for larger $\varepsilon$,
 we need a positive $\eta$, which is realized
 for a negative $\cos \theta$.
As mentioned above, the contribution of $\xi$ is important 
to realize a large negative $\alpha_s$.
For this purpose, $\xi$ must be positive.
Then, $\tan \theta$ must be negative.
For $\varepsilon \sim 0.01$ and $\eta \sim 0.01$ 
as well as $A\sim x$, we obtain $2\xi \sim 0.02$ 
by $\tan \theta \sim {\cal O}(10)$.
We show some examples in Table~\ref{tab:double} as well as Fig.~\ref{Fig:asrt}, 
where $N$ denotes the e-folding generated through 
our model at the first stage of inflation.

\begin{table}[h]
\begin{tabular}{|cccc|cccc|}\hline
~~~~$x$ ~~~~&~~~~  $A$ ~~~~&~~~~ $\phi$ ~~~~&~~~~ $\theta$ ~~~~
& ~~~~$n_s$ ~~~~&~~~~ $r_T$~~~~ &~~~~ $N$ ~~~~&~~~~ $\alpha_s$~~~~ \\
\hline \hline
0.2 & 0.1 & 7.8 & $4 \pi/9$ & 0.951 & 0.084 & 31 & $-0.022$ \\
0.3 & 0.2 & 6.3 & $17 \pi / 36$ & 0.951 & 0.098 & 21  & $-0.025$ \\
0.3 & 0.3 & 4.4 & $5 \pi / 9$  & 0.953 & 0.187 & 12 & $-0.021$ \\
0.4 & 0.4& 3.9 & $19 \pi / 36$ & 0.953 & 0.17 & 11 & $-0.030$ \\
0.4 & 0.5 & 4.0 & $5 \pi/9$   & 0.955 & 0.20 & 11 & $-0.021$ \\
\hline
\end{tabular}
\caption{Results in the double inflation scenario}
\label{tab:double}
\end{table}

%
%

As mentioned above, we assume that 
the second inflation follows after the above monodromy inflation occurs
 such that the total number of e-folds becomes $50$ or more.
In principle, any inflation model can be allowed 
and there is no constraint at present.
Hence, we do not analyze its model-dependent detail here.

\section{Summary}

We have studied axion monodromy inflation with 
a sinusoidal correction term.
We find the sinusoidal correction term is important in estimating
 inflationary observables; in fact,
 $n_s$ and $r_T$ are significantly affected by the correction.
We have shown that a larger tensor-to-scalar ratio compatible with the recent BICEP2 results
 can be obtained in both the single and double inflation scenarios. 
A large negative running of the spectral index can be realized
 in the double inflation scenario, whose first stage is driven
 by our monodromy inflation model with sinusoidal correction term.


\section*{Acknowledgments}
This work was supported in part by the Grant-in-Aid for Scientific Research 
No.~25400252 (T. K.) and on Innovative Areas No.~26105514 (O. S.)
 from the Ministry of Education, Culture, Sports, Science, and Technology in Japan. 
%



\end{document}